\documentclass[conference]{IEEEtran}
\IEEEoverridecommandlockouts
\usepackage{cite}
\usepackage{amsmath,amssymb,amsfonts}
\usepackage{algorithmic}
\usepackage{graphicx}
\usepackage{textcomp}
\usepackage{epsfig}
\usepackage{multirow}
\usepackage{float}
\usepackage{subfigure}
\usepackage{booktabs}
\def\BibTeX{{\rm B\kern-.05em{\sc i\kern-.025em b}\kern-.08em
		T\kern-.1667em\lower.7ex\hbox{E}\kern-.125emX}}
\begin{document}
	
	\title{Chain-Net: Learning Deep Model for Modulation Classification Under Synthetic Channel Impairment\\
		\thanks{This research was financially supported by National Research Foundation of Korea (NRF) through Creativity Challenge Research-based Project (2019R1I1A1A01063781), and in part by the Priority Research Centers Program through the NRF funded by the Ministry of Education, Science and Technology (2018R1A6A1A03024003).}
	}
	
	\author{Thien Huynh-The$^{\ast }$, Van-Sang Doan$^{\ast }$, Cam-Hao Hua$^{\dagger}$, Quoc-Viet Pham$^{\ddagger }$, and Dong-Seong Kim$^{\ast }$\\
		$^{\ast }$ICT Convergence Research Center, Kumoh National Institute of Technology, Korea\\
		$^{\dagger}$Department of Computer Science and Engineering, Kyung Hee University, Korea\\
		$^{\ddagger}$Research Institute of Computer, Information and Communication, Pusan National University, Korea\\
		 Email: \{thienht,vansang.doan,dskim\}@kumoh.ac.kr,~hao.hua@oslab.khu.ac.kr,~vietpq@pusan.ac.kr}

	\maketitle

\begin{abstract}
Modulation classification, an intermediate process between signal detection and demodulation in a physical layer, is now attracting more interest to the cognitive radio field, wherein the performance is powered by artificial intelligence algorithms.
However, most existing conventional approaches pose the obstacle of effectively learning weakly discriminative modulation patterns.
This paper proposes a robust modulation classification method by taking advantage of deep learning to capture the meaningful information of modulation signal at multi-scale feature representations.
To this end, a novel architecture of convolutional neural network, namely Chain-Net, is developed with various asymmetric kernels organized in two processing flows and associated via depth-wise concatenation and element-wise addition for optimizing feature utilization.
The network is evaluated on a big dataset of 14 challenging modulation formats, including analog and high-order digital techniques.
The simulation results demonstrate that Chain-Net robustly classifies the modulation of radio signals suffering from a synthetic channel deterioration and further performs better than other deep networks.
\end{abstract}


\section{Introduction}
Automatic modulation classification, a current attractive topic in artificial intelligence (AI)-powered wireless communication, plays the central function of intelligent spectrum management, in which the modulation format of an incoming signal is precisely identified at the receiver based on analyzing radio characteristics~\cite{intro01,intro01-1,intro01-2}.
Typically, automatic modulation classification methods can be categorized into two groups of likelihood-based (LB) and feature-based (FB).
LB approaches return the output modulation by maximizing the probability of a received signal mostly associated with a certain modulation. 
This procedure is done with the model parameters estimated by the expectation/conditional maximization (ECM) algorithm~\cite{intro02}.
Despite promisingly resulting in an optimal accuracy in the ideal scenario (where the information of signal and transmission channel is completely acknowledged), LB approaches face two primary obstacles of being failure with many continuous phase modulations and consuming expensively computational resource~\cite{intro03}.
Compared with LB methods, FB approaches have several advantages of easier implementation, lower complexity, and stronger robustness under a synthetic channel deterioration by exploiting advanced feature engineering techniques and machine learning (ML) algorithms~\cite{intro04}.
Following the general ML workflow for classification task, most current FB methods extract handcrafted features (such as high-order statistics with moments and cumulants, cross-correlation characteristics, and cyclostationary properties) for model learning with different traditional classification algorithms.
The principle drawback of FB approaches involves the necessity of feature engineering expertise and the less capability of dealing with big data~\cite{intro05}.
Due to the poorly discriminative ability of shallow features, current conventional methods cannot classify the inter-group modulations and the intra-group high-order modulations properly under heavy channel impairments.

Recently, deep learning (DL)~\cite{intro06} has been studied successfully in various domains, from computer vision~\cite{intro07,intro07-1,intro07-2} to biomedical engineering~\cite{intro08,intro08-1}, thanks to the advantage of non-linear formulation via activation functions embedded in multiple hidden layers.
Accordingly, the intrinsic information in high-dimensional unstructured data is exhaustively learned at multi-scale feature maps.
Some DL models have been introduced for modulation classification, but without network architecture optimization, their performance cannot satisfy modern communication services~\cite{intro09}.
In this paper, we introduce a novel convolutional neural network (CNN), namely Chain-Net, for automatic modulation classification. 
The network architecture, inspired by the chain shape, is structured to optimize the utilization of multi-scale feature maps.
Leveraging various one-dimensional (1-D) asymmetric convolution kernels for depth-wise concatenation and element-wise addition exhaustively gains the cross-component correlation within a sample and the sample-wise relation in a radio signal.
Besides computational complexity, vanishing gradient and overfitting issues are taken into account in Chain-Net.
For performance evaluation, we further contribute a challenging dataset which covers up to 14 modulation fashions, including analog and digital modulations with high-order formats, under a synthetic impairment of Rayleigh multipath fading channel and additive noise. 
Based on simulation results, Chain-Net significantly outperforms other existing CNN-based modulation classification approaches.

\section{Related Work}
Xie \textit{et al.}~\cite{work01} introduced a DL approach to improve the accuracy of recognizing six digital modulations, in which the handcrafted high order cumulants are calculated for model learning with a deep neural network (DNN).
The network architecture is naively designed with one input layer, two hidden layers (wherein the rectified linear unit function is embedded for non-linear transformation), and one output layer.
Due to the small-scale learning capacity of DNN and the poorly discriminative characteristic of cumulant feature, this approach is almost tackled by advanced high-order modulations.
Hu \textit{et al.}~\cite{work02} developed a robust modulation classification method by studying the advantage of long short-term memory (LSTM) network, an advanced-architecture variant of recurrent neural network (RNN) for learning long-term dependent features from the array of in-phase and quadrature (I/Q) samples.
The LSTM network is principally carved out by three stacked-LSTM layers and four fully connected (fc) layers.
The learning model is evaluated with four less challenging modulation formats (e.g., BPSK, QPSK, 8-PSK, and 16-QAM) in a positive range of signal-to-noise ratio (SNR).
Despite being better than traditional ML-based and baseline RNN-based approaches, the method is not suitable for the consideration of more modulation formats.

Recently, CNNs have been intensively studied for learning modulation patterns from I/Q data and also image-based data, such as spectrogram and constellation diagram.
Meng \textit{et al.}~\cite{work03} designed a CNN architecture with several conventional stacks (where each stack involves a convolutional layer, a ReLU layer, and a max pooling layer) connected with two fc layers for seven-modulation classification. 
By associating with the SNR information, the network can achieve remarkable performance in terms of classification rate, however, a huge number of parameters in two fully connected layer encumbers the network more heavily.
In another work, Zeng \textit{et al.}~\cite{work04} built a conventional CNN for learning visual representational features of spectrogram images which can be obtained via calculating the squared magnitude of the short-time discrete Fourier transform of a modulated signal.
The network involving four convolutional layers can relatively cope with up to 11 modulation formats, including digital and analog styles, but its unsophisticated-designed architecture cannot seize the intrinsic correlation between samples besides the lack of handling overfitting and vanishing gradient problems.

In~\cite{work05}, Huang \textit{et al.} generated the regular constellation diagram of modulation signal for learning a compact-sized CNN, in which the multi-scale visual features are useful for intra-class discrimination. 
The image-based classification model is experimented for five cushy modulation fashions (including BPSK, QPSK, 8-PSK, 16-QAM, and 64-QAM) and consequently obtains some promising results, however, this strategy unavoidably faces some image-processing-related issues (for example, pepper noise and image quality degradation).
To overcome the aforementioned problem, Wang \textit{et al.}~\cite{work06} introduced a hierarchical learning model which is organized by two CNNs for concurrently taking into account both the I/Q sample and the constellation diagram of an incoming signal.
Even though this approach induces some interesting accuracy results, its costly computational complexity can become a practical challenge.
Regarding the literature review of state-of-the-art works, it is observed that most of the deep networks with naive architecture cannot optimize feature manipulation to not only deal with numerous high-order modulation formats but also improve the classification rate under various channel impairments.

\section{Deep Convolutional Neural Network for Modulation Classification}
\subsection{Problem Definition}
In modern wireless communication systems, many advanced digital and analog modulations have been deployed to optimize the spectrum utilization. 
The characteristics (including amplitude, frequency, phase, and amplitude-phase hybrid) of carrier signal (referred to as high-frequency periodic waveform) are changed by a pre-defined modulation scheme for encoding the meaningful data for communication.
Fundamentally, automatic modulation classification aims to identify the most appropriate modulation format of an incoming modulated signal via a trainable classifier embedded at the receiver.
With regard to several conventional approaches, machine learning algorithms have been investigated for learning the radio characteristics, in the time and frequency domains, of I/Q samples, in which some sophisticated signal processing and feature engineering techniques are jointly taken into account for fine-tuning classification performance.
Let $x(t)$ be the transmission signal bearing a synthetic impairment which includes the channel effect $h(t)$ and the additive noise $n(t)$, the modulation signal $y(t)$ acquired at the receiver is generally written as follows:
\begin{equation}
y(t)=x(t) \ast h(t)+n(t).
\label{eqn:01}
\end{equation}
Regarding a regular supervised learning-based approach, a classifier is able to perceive various modulation patterns during the training process with a given dataset.
Accordingly, the trained classifier can predict the modulation fashion of $x(t)$ based on the inference procedure of $y(t)$.

\begin{figure}[!t]
	\centering
	\subfigure[]
	{	\includegraphics[width=2.8in]{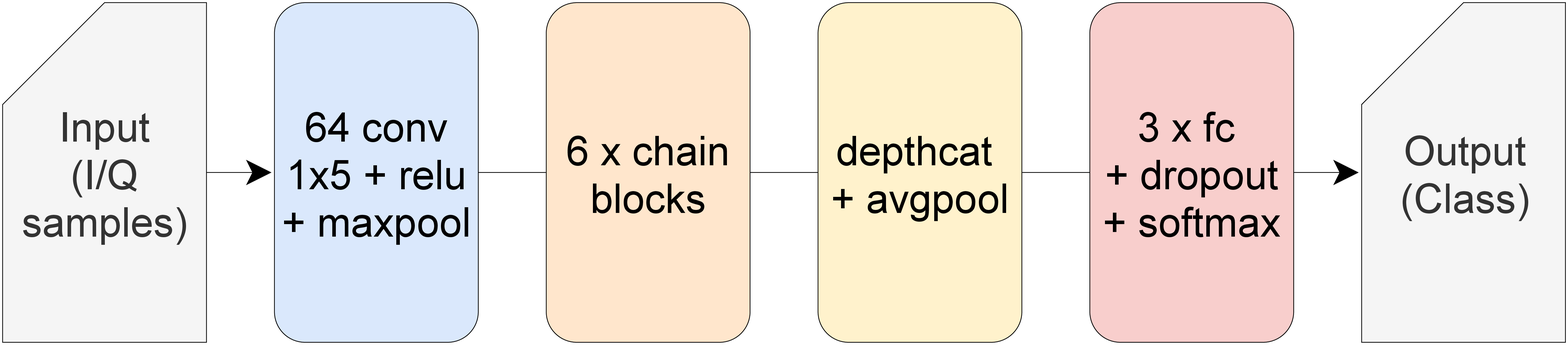}
		\label{architecture}
	}
	\subfigure[]
	{	\includegraphics[width=2.8in]{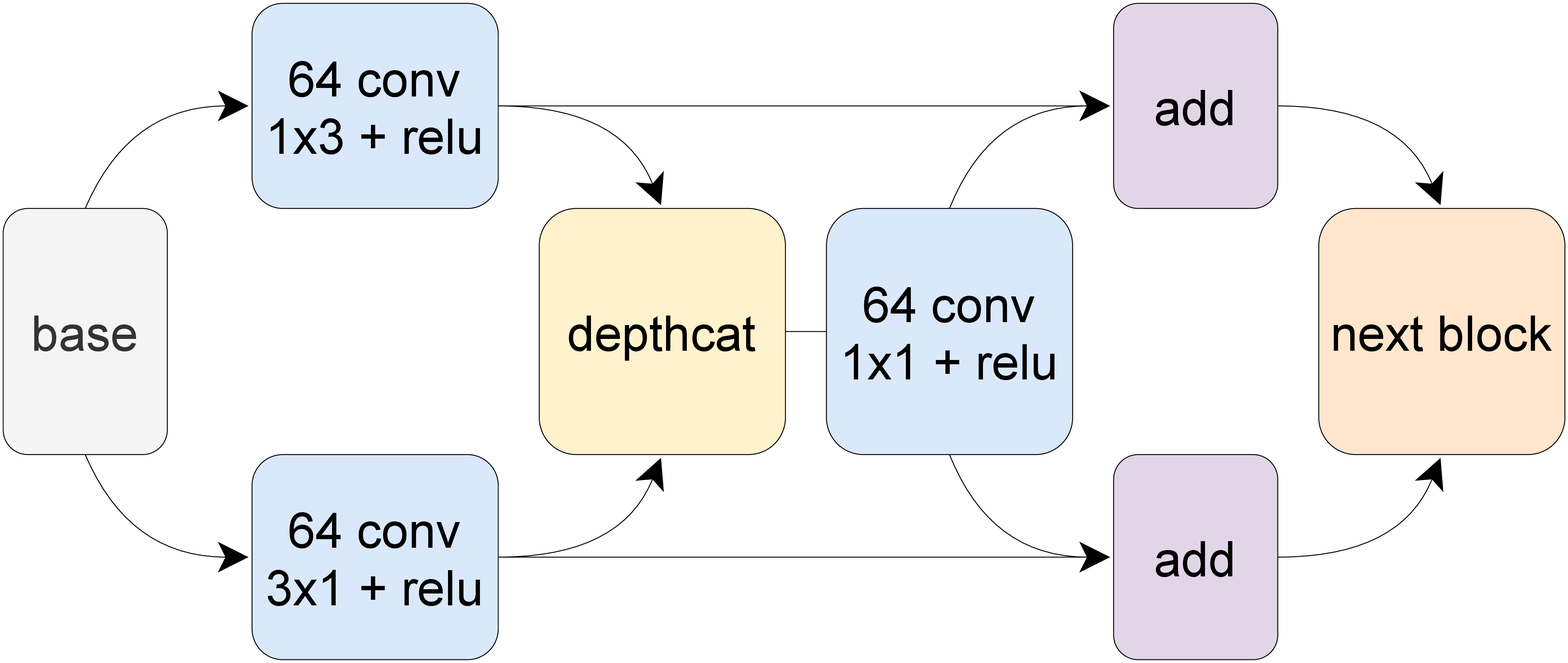}
		\label{chainblock}
	}
	\caption{Chain-Net: (a) the overall architecture principally involves six convolutional blocks that are organized in a cascade for deep feature extraction and (b) each block is specified by two asymmetric (including horizontal $1 \times 3$ and verical $3 \times 1$ kernels) convolutional flows typically incorporated via depth-wise concatenation and element-wise addition.}
	\label{network}
\end{figure}

\subsection{Chain-Net Architecture}

In this section, we study an automatic modulation classification method based on taking advantage of CNNs to learn multiple scaling features.
As the primary contribution, a deep CNN, namely Chain-Net, is comprehensively designed to satisfy the high performance of communication services in terms of high accuracy and low latency.
It is worth noting that the input of network is I/Q samples which are partitioned into a single signal frame with a pre-defined length $\ell = 1024$ (referred to as the number of I/Q samples).
As shown in Fig.~\ref{architecture}, the network architecture is mainly constituted by six convolutional blocks for extracting representational features at different resolutions.
Chain-Net is initialized with an input layer configured by the size of $2 \times 1024$ for being compatible with I/Q data.
A basic processing stack, which consists of a convolutional layer ($\texttt{conv}$) with 64 the kernels of size $1 \times 5$, an activation layer with the rectified linear unit function ($\texttt{ReLU}$) and a max pooling ($\texttt{maxpool}$) layer with the pool of size $1 \times 2$, is specified at the beginning of network.
Being the principal operation in CNNs, convolution is the dot product of the weights $w_k$ of a given kernel (so-called filter) and the elements $a_k$ of an input map that locate inside a local region (aka receptive field), where the depth of kernel is properly identical to that of the input map.
Afterwards, an optional scalar bias $b$ is further added to the result of convolution for a better linear transformation and thus the output of a convolutional layer at any spatial coordinate $\left ( i,j \right )$ can be expressed as follows
\begin{equation}
u_{i,j}={\sum_{k}{w_k a_k}}+b.
\label{eqn:02}
\end{equation}
By passing scalar $u$ through an activation ReLU function $g$, a feature map (aka activation map) is obtained by scaling the negative input element to zero as follows
\begin{equation}
v_{i,j}=g\left ( u_{i,j} \right )=\begin{cases}
u_{i,j} & \text{ if }  u_{i,j}\geq 0, \\ 
0 & \text{ if }  u_{i,j} < 0.
\end{cases}
\label{eqn:03}
\end{equation} 
According the similar operating principle of the convolutional layer, the max pooling layer responds the maximum value within a pre-defined region but there is no parameter learning.
With the objective of quickly reducing the dimensionality of feature maps, the stride specified for the convolutional layer and the max pooling layer in the block is $\left ( 1,2 \right )$ to reduce the horizontal size of the input maps by four times.

\begin{figure}[!t]
	\centering
	\subfigure[]
	{	\includegraphics[width=1.7in]{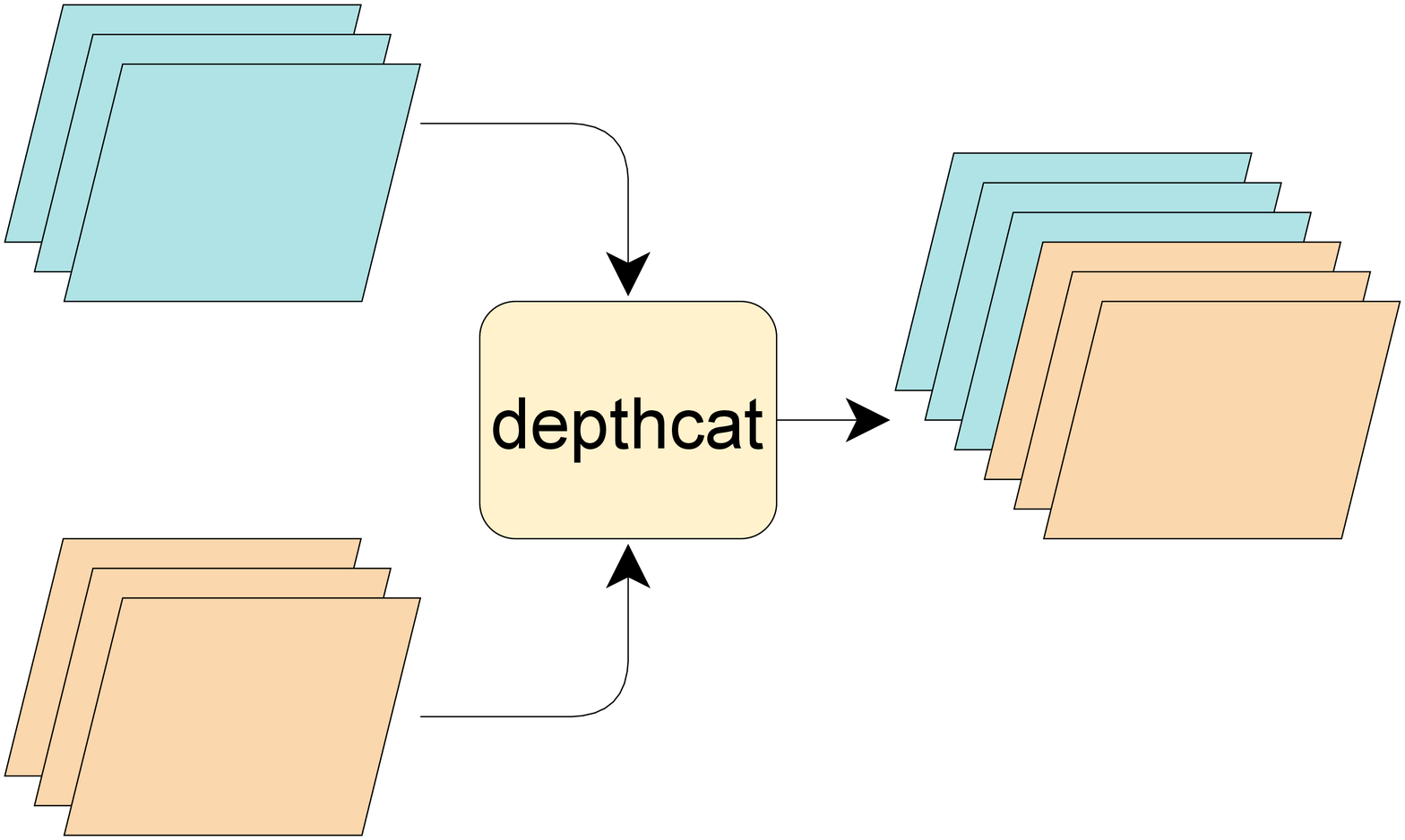}
		\label{depthcat}
	}
	\subfigure[]
	{	\includegraphics[width=1.4in]{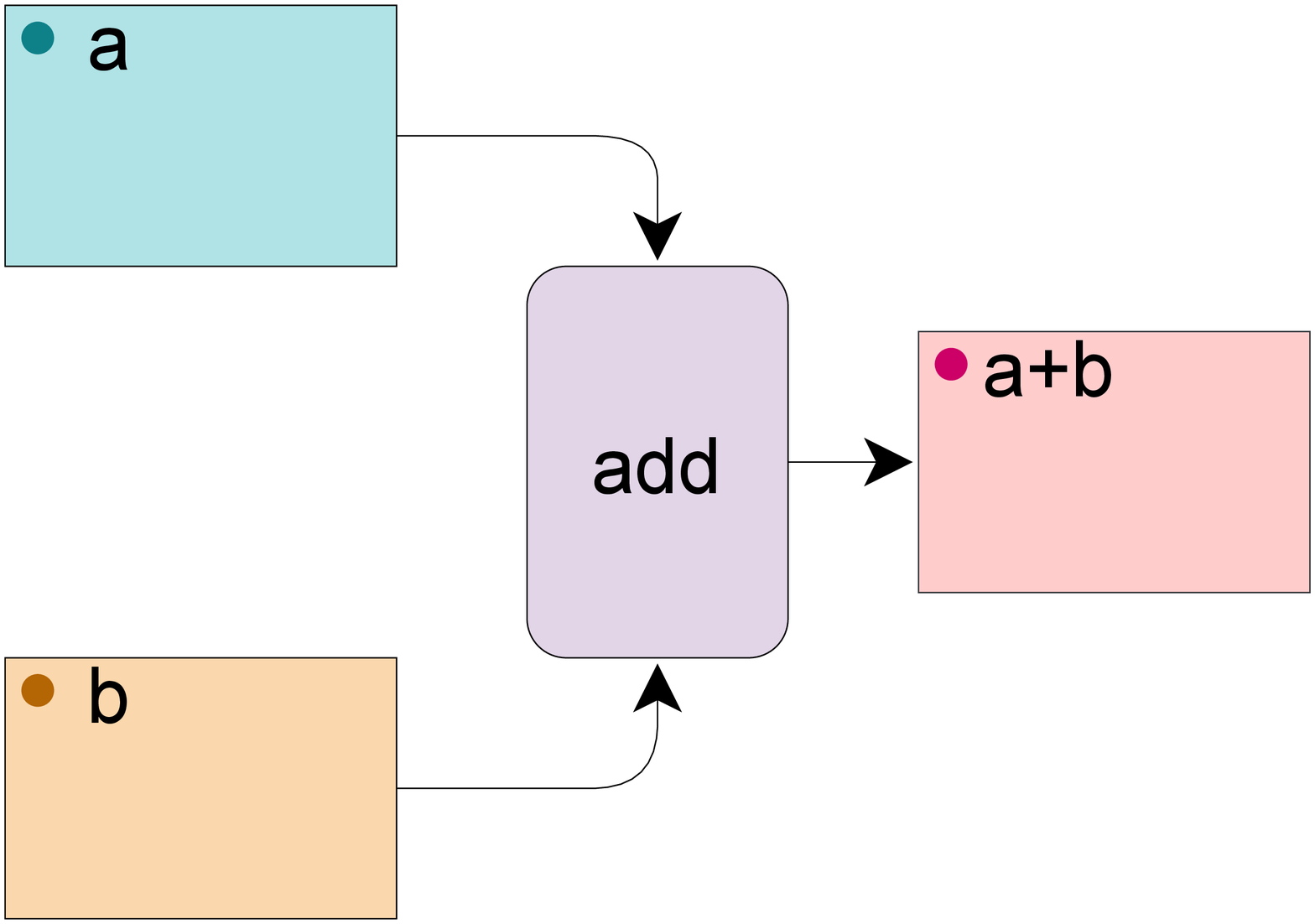}
		\label{add}
	}
	\caption{Illustration of layer operation: (a) depth-wise concatenation layer stacks the feature maps of two inputs along the depth dimension and (b) addition layer calculates the sum of two input maps following the element-wise manner.}
	\label{operation}
\end{figure}


\begin{table}
	\centering
	\footnotesize
	\setlength{\tabcolsep}{2pt}
	\caption{Detailed configuration of network architecture.}
	\begin{tabular}{|l|c|c|}
		\hline
		\textbf{Component} & \textbf{Output} & \textbf{Detailed description} \\
		\hline
		$\mathtt{input}$ 	& $2 \times 1024 \times 1$ 	& $1024~\mathtt{I/Q~samples}$ \\ \hline
		$1 \times\mathtt{stack}$ & $2 \times 256 \times 64$ & $1 \times \left \{ \begin{matrix}
		64~\mathtt{conv}~1\times5,\mathtt{~stride}~\left (1,2  \right )\\ 
		\mathtt{ReLu~activation~function}\\ 
		\mathtt{max~pool~1\times2},\mathtt{~stride}~\left (1,2  \right )
		\end{matrix} \right \}$ \\ \hline
		$6 \times~\mathtt{block}$ & $2 \times 4 \times 64^\ast$ & $6 \times \left \{ \begin{matrix}
		64~\mathtt{conv}~1\times3,\mathtt{~stride}~\left (1,2  \right )\\ 
		64~\mathtt{conv}~3\times1,\mathtt{~stride}~\left (1,2  \right )\\ 
		64~\mathtt{conv}~1\times1,\mathtt{~stride}~\left (1,1  \right )\\ 
		3\times\mathtt{ReLu~activation~function}\\ 
		1\times\mathtt{depth-wise~concatenation}\\ 
		2\times\mathtt{element-wise~addition}
		\end{matrix} \right \}$ \\ \hline
		$\mathtt{depthcat}$ & $2 \times 4 \times 128$ & $\mathtt{depth-wise~concatenation}$\\ 
		$\mathtt{avgpool}$ & $1 \times 1 \times 128$ & $\mathtt{average~pool~2\times4}$\\ \hline
		$3 \times \mathtt{fc}$ & $1 \times 1 \times 14$ &  $\left \{ \begin{matrix}
		\mathtt{fc1:~128~nodes}\\ 
		\mathtt{fc2:~128~nodes}\\ 
		\mathtt{fc3:~14~nodes}
		\end{matrix} \right \}$ \\ \hline
		\addlinespace[0.1cm]
		\multicolumn{3}{l}{$\ast$ This refers to as the volume size of the feature resulted in each flow.} \\
	\end{tabular}
	\label{configuration}
\end{table}

Six convolutional blocks (called chain block) are next arranged in the network architecture, which aims to optimally learn representational features at multi-scale maps.
The chain block, where its inside structure as shown in Fig.~\ref{chainblock} is inspired from the shape of catenary, mainly involves two convolutional flows in parallel.
For details, each flow has a convolutional layer which is specified by 1-D asymmetric kernels, either the horizontal kernel of size $1 \times 3$ or the vertical kernel of size $3 \times 1$.
For being compatible with the convolution of vertical kernels, zero padding is supplemented along the border of input map correspondingly.
With the horizontal kernels, the network can calculate the locally temporal relation between I/Q samples, meanwhile, the cross-component correlation of each sample is seized by the vertical kernels.
Compared with the 2-D kernel of size $3 \times 3$, the parallel arrangement of two 1-D asymmetric kernels in the chain block is cheaper while gaining a same learning efficiency~\cite{method01} confidently.
The outputs from two corresponding flows, denoted $y_{\texttt {flow-} 1\times 3}$ and $y_{\texttt {flow-} 3\times 1}$, are accumulated by a depth-wise concatenation layers ($\texttt{depthcat}$) as follows
\begin{equation}
y_{\texttt {depthcat}}=\texttt{concat} \left \{ y_{\texttt {flow-} 1\times 3}, y_{\texttt {flow-} 3\times 1} \right \}.
\label{eqn:04}
\end{equation} 
As the operation illustrated in Fig.~\ref{depthcat}, two inputs having the same height and width are concatenated along the depth dimension, hence the depth size of output volume is the sum of those of two inputs.
The informative features from two convolutional flows are incorporated at once in $y_{\texttt {depthcat}}$ that is favorable to enrich the discrimination of feature and further mitigate the vanishing gradient caused by activation function in each flow.
To this end, we deploy an addition layer ($\texttt{add}$), where all inputs must have the same dimension, to combine the output of $\texttt{depthcat}$ with each processing flow via an element-wise addition operation, however, the connection from $\texttt{depthcat}$ to $\texttt{add}$ should be done over an intermediate unit convolution layer with the kernel of size $1 \times 1$ to re-scale depth dimension.
This procedure can be expressed as follows
\begin{equation}
y_{\texttt{add}} = y_{\texttt {flow}} + \texttt{conv}_{1\times 1}\left ( y_{\texttt {depthcat}} \right ).
\label{eqn:05}
\end{equation} 
With this mechanism, the information identity of each flow is synthesized with less gradient degradation.
Additionally, the discriminative information seized in multi-scale representational feature maps along multiple chain blocks is enhanced for being more multifarious.

At the end of network, we assemble the feature maps returned by two flows of the last chain block by a depth-wise concatenation layer.
The architecture is finalized with a global average pooling layer ($\texttt{avgpool}$), three fully connected layers ($\texttt{fc}$) (where the number of neurons, denoted $C$, of the last one is identical to the number of modulation formats in a given dataset), a dropout layer with the dropping ratio of $0.5$ for handling overfitting issue, and a softmax layer.
The detailed configurations of network architecture are summarized in Table~\ref{configuration}.
With respect to multi-class classification tasks, the softmax layer, which must follow the final $\texttt{fc}$ layer, is able to respond the probability of all classes at once using a softmax function (also known as the normalized exponential) as follows
\begin{equation}
y_c\left ( x \right )=\frac{e^{\alpha _c\left ( x \right )}}{\sum_{l=1}^{C}{e^{\alpha _l\left ( x \right )}}} ~~~\textrm{for}~c=1\dots C,
\label{eqn:06}
\end{equation} 
where $0\leq y_c\leq 1$ and $\sum_{l=1}^{C}y_c = 1$.
In the training procedure, the cross entropy loss is estimated between the target class (aka ground-truth class) and the output classed resulted by feedforward estimation as follows
\begin{equation}
\mathcal{L} = -\sum_{p = 1} ^ {N}\sum_{q = 1}^{C} \upsilon _{pq} \mathtt {ln} \left( {\nu}_{pq} \right),
\label{eqn:07}
\end{equation} 
where $N$ is the number of training signals, $\upsilon _{pq}$ denotes that the $p^{th}$ signal belongs to the $q^{th}$ modulation format as the truth, and ${\nu}_{pq}$ remarks the prediction class $q$ for the signal  $p$ given by the network. 
Chain-Net is trained from scratch using randomly initialized weights in 100 epochs with the stochastic gradient descent algorithm for optimally updating parameters, the initial learning rate of 0.01, and the mini-batch size of 256.

\section{Performance Evaluation}

\subsection{Dataset Generation Approach}

For the performance evaluation of the proposed DL model for modulation classification, we generate a new dataset that consists of 14 different formats of analog (e.g., AM$-$DSB$-$WC, AM$-$DSB$-$SC, AM$-$SSB$-$WC, AM$-$SSB$-$SC, FM) and digital (e.g., 16PAM, 16QAM, 32QAM, 64QAM, 128QAM, 16APSK, 32APSK, 64APSK, 128APSK) modulation.
Regarding propagation channel scenario, we follow a typical multipath fading model, namely Extended Pedestrian A (EPA) model~\cite{EPA01} and~\cite{EPA02}, to synthesize the modulation signals under additive white Gaussian noise (AWGN) at SNR varying from -20 dB to +20 dB with a step size of 2 dB.
The properties of Rayleigh channel over EPA is specified as follows: the propagation path delays in range $\left \{0, 30, 70, 90, 110, 190, 410  \right \}$ ns, the average path gains in range $\left \{0, -1.0, -2.0, -3.0, -8.0, -17.2, -20.8  \right \}$ dB, and the maximum Doppler shift of 10 Hz. 
For analog modulations, we set the frequency of carrier signal $f_c = 5$ MHz and the sampling frequency $f_s = 20f_c$ (where the envelope signal is an audio signal).
By generating 4000 1024-sample signals per modulation formats per SNR, the dataset has 1,176,000 signals in total.
Compared with many previous works~\cite{work01, work02, work03, work04, work05}, which have investigated a few effortless modulation fashions without multipath propagation, this study takes into account numerous challenging high-order digital modulation formats under the aforementioned synthetic channel impairment.

\begin{figure}[!t]
	\centering
	\includegraphics[width=8cm]{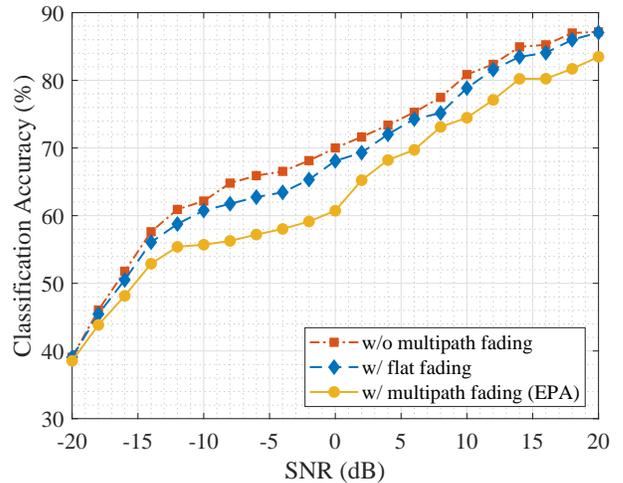}
	\caption{Classification accuracy of Chain-Net under different channel impairment scenarios.}
	\label{fig_robustness}
\end{figure}

\subsection{Model Robustness} 

\begin{figure*}[!t]
	\centering
	\subfigure[]
	{	\includegraphics[width=3in]{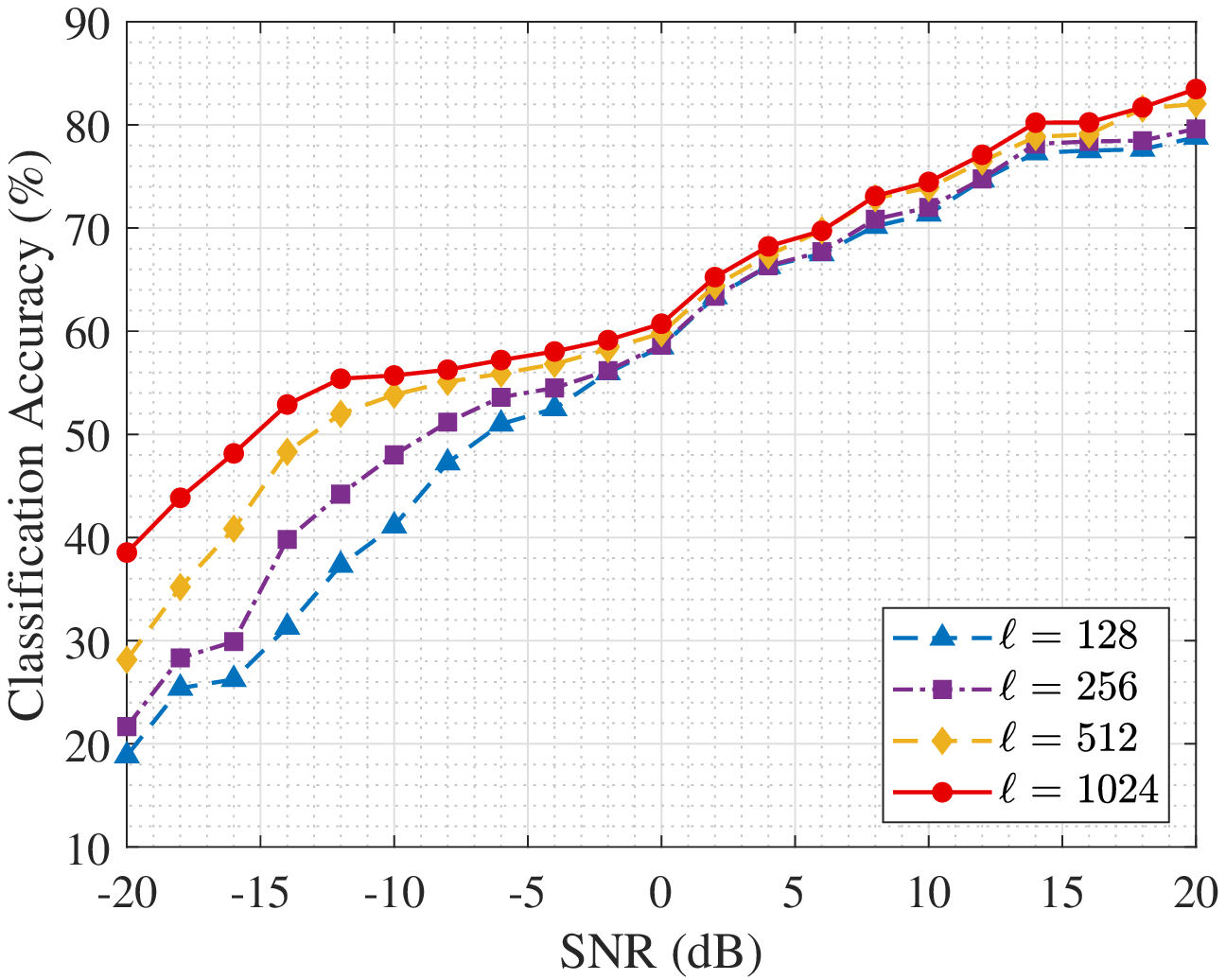}
		\label{fig_signallength}
	}
	\subfigure[]
	{	\includegraphics[width=3in]{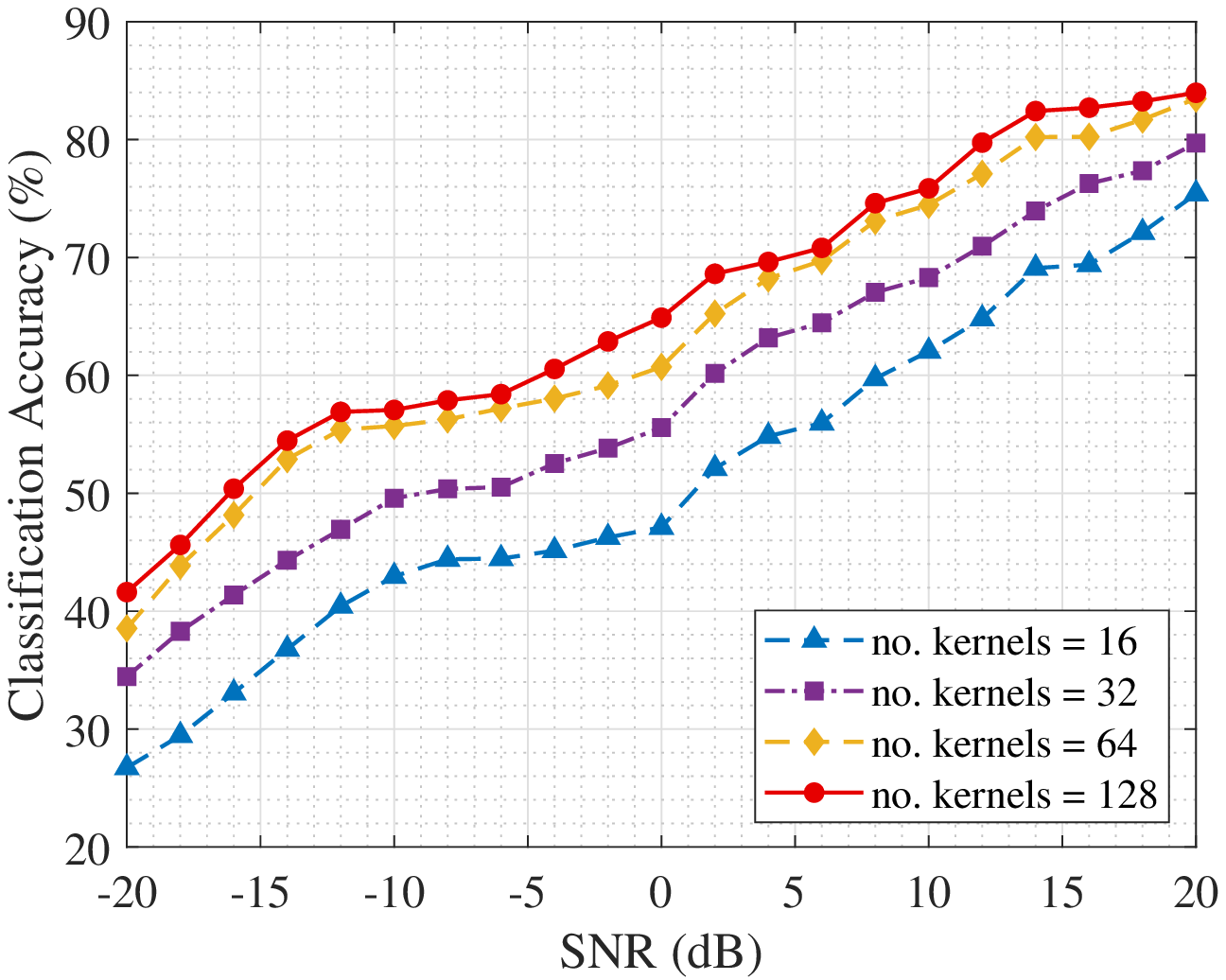}
		\label{fig_numberkernel}
	}
	\caption{Benchmark of performance sensitivity under different hyper-parameter values: (a) the signal length $\ell$ and  (b) the number of kernels specified in convolutional layers.}
	\label{fig_sensitivity}
\end{figure*}


In this first experiment, we report the numerical results of 14-modulation classification in Fig.~\ref{fig_robustness}, where Chain-Net achieves the overall accuracy (for all examined SNR levels) of $63.78\%$ (approximately $83.5\%$ at $+20$ dB SNR) under the synthetic channel impairment of multipath Rayleigh fading channel over EPA.
To investigate the robustness of deep model, we further measure the performance in two common scenarios: without (w/o) multipath fading channel and with (w/) flat fading channel.
It is observed that the accuracy increases along the increment of SNR in general.
The harmful effect of flat fading channel on to the modulation signals is minor with the overall accuracy reduction of $1.65\%$, decreasing from $69.45\%$ (w/o multipath fading) to $67.80\%$ (w/ flat fading).
Meanwhile, the performance significantly decreases by about $5.67\%$ under a strong channel deterioration caused by multipath Rayleigh fading, where the channel is frequency-selective nature with the negative path gains and the Dopper shift.
Due to the critical modification of radio signal characteristics, including amplitude and phase caused by the multipath fading, many high-order modulations, such as 64QAM, 64 APSK, 128QAM, and 128APSK, are misclassified considerably.
From $-10$ dB to $0$ dB SNR, the multipath fading knocks the network performance down approximately $8.47\%$ as average lower accuracy.

\subsection{Performance Sensitivity}

The second experiment aims to analyze the performance sensitivity of the proposed CNN-based modulation classification method, in which two hyper-parameters of Chain-Net are investigated for accuracy measurement.
Concretely, we vary the signal length $\ell$ (aka the number of I/Q samples) in range $\left \{ 128,256,512,1024 \right \}$ and the number of kernels specified in convolutional layers in range $\left \{ 16,32,64,128 \right \}$, where the quantitative results are presented in Fig.~\ref{fig_sensitivity}.
With $\ell = 1024$, Chain-Net achieves the best performance at multiple SNR levels, for instance, $74.47\%$ at 10 dB SNR and up to $83.47\%$ at 20 dB SNR.
Interestingly, by increasing the signal length from $512$ to $1024$, the classification accuracy is strongly improved by around $4.08\%$ higher on average at $\textrm{SNR}\leq 0$ dB and slightly enhanced by approximately $0.72\%$ at high SNR ($\geq 2$ dB) as shown in Fig~\ref{fig_signallength}.
Consequently, it is recognized that significant accuracy improvement is mostly obtained at low SNRs for each time of doubling the number of I/Q samples for classification.
For example, at $-10$ dB SNR, Chain-Net successfully raises the accuracy by $6.88\%$, $5.83\%$, and $1.87\%$ for successive times of doubling the signal length $\ell = 128$.
Obviously, Chain-Net can reach a higher modulation classification with a longer signal length (i.e., more I/Q samples in a partitioned signal) because more meaningful intrinsic information of radio characteristics can be explicitly learned via a deep architecture. 
However, the system computational complexity is accordingly more expensive for traversing the receptive field of convolutional kernel onto a larger spatial size of feature maps, besides the costly memory assumption of data repository.

With regard to the number of kernels specified in convolutional layers, the assessment results are plotted in Fig.~\ref{fig_numberkernel}.
It is realized that the classification performance of Chain-Net is properly enhanced by increasing the number of 1-D asymmetric and unit kernels in chain blocks.
The overall accuracy rate is significantly improved by approximately $6.99\%$ and $5.72\%$ when increasing the kernel number from 16 to 32 and from 32 to 64, respectively.
In details, Chain-Net achieves the accuracy rate of $74.47\%$ with 64 kernels at $+10$ dB SNR, which is greater than that with 32 kernels by $6.16\%$.
By deploying more kernels, Chain-Net is able to capture more cross-component correlations of each sample and temporal relations between different consecutive samples via more vertical $1\times3$ and horizontal $3\times1$ kernels, respectively.
Notably, doubling the number of kernels from 64 to 128 gains a minor accuracy enhancement of around $2.04\%$, but the network efficiency, denoted as the ratio of accuracy improvement over system complexity growth, reduces in non-linearity.
With large kernel numbers, the engendered tiny accuracy improvement is inappropriate to the rapidly increasing network capacity (quantified by the number of network parameters including weights and biases).
Therefore, regarding 64 convolution kernels configured in layers, Chain-Net shows a reasonable trade-off between accuracy and inference speed (or referred to as processing latency) for high-reliability services promisingly.

\subsection{Method Comparison}
\begin{figure}[!t]
	\centering
	\includegraphics[width=8cm]{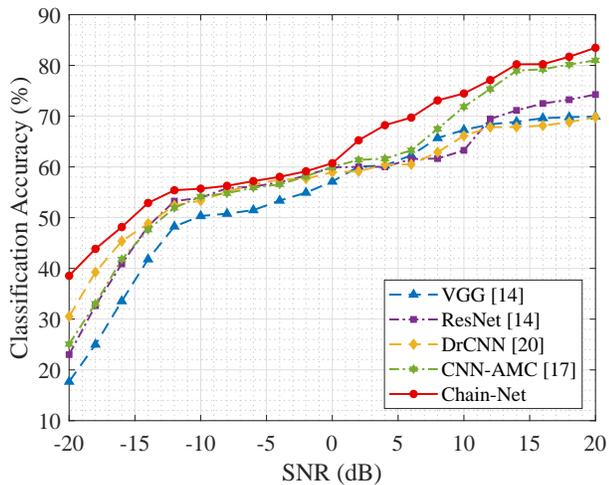}
	\caption{Method comparison in terms of 14-modulation classification accuracy.}
	\label{fig_comparison}
\end{figure}

The last experiment compares the classification accuracy between Chain-Net with other state-of-the-art CNN models for modulation classification, including VGG~\cite{intro09}, ResNet~\cite{intro09}, CNN-AMC~\cite{work03}, and DrCNN~\cite{work06}, under the multipath fading condition.
From the results plotted in Fig.~\ref{fig_comparison}, VGG with the straightforward architecture specified by seven convolutional layers presents the worst performance.
By taking advantage of residual connection to maintain the information identity throughout the network architecture, ResNet is better VGG at low SNRs while utilizing fewer kernels.
Different from VGG and ResNet which deploy 1-D kernels of small size $1 \times 3$, DrCNN specifies different kernels of larger size of $2 \times 8$ and $1 \times 16$ to capture more relevant information in a bigger receptive field.
As the benefit of configuring more hidden nodes in two fully connected layers, CNN-AMC encouragingly improves classification accuracy at high SNRs ($\geq 6$ dB), but the computational complexity rapidly increases along the growth of network capacity.
Compared with these CNNs, Chain-Net achieves the best performance with the sophisticated structure of chain block, where the informative features are enhanced with depth-wise concatenation and element-wise addition.

\section{Conclusion}

In this paper, we have introduced an efficient deep network, namely Chain-Net, for modulation classification, in which the architecture is specified by multiple convolutional blocks to exhaustively extract more relevant information.
By deploying 1-D asymmetric convolution kernels, the designed network is capable of exposing the cross-component correlation within a sample and the sample-wise relation of a radio signal.
Through the performance benchmark on the dataset of 14 challenging modulation formats, Chain-Net achieves the classification rate of approximately $83.5\%$ at $+20$ dB SNR under a multipath Rayleigh fading channel, in which several intensive simulations are comprehensively provided to investigate performance sensitivity under various hyper-parameter configurations.
Remarkably, with a well-designed structure, Chain-Net outperforms many state-of-the-art deep models for modulation classification in terms of accuracy while maintaining a cheap network size.
Future work will focus on upgrading the network architecture to yield a better classification performance with more modulation formats.
%



\begin{thebibliography}{00}
\bibitem{intro01} T. Huynh-The, C. Hua, Q. Pham and D. Kim, ``MCNet: An Efficient CNN Architecture for Robust Automatic Modulation Classification,'' \textit{IEEE Commun. Lett.}, vol. 24, no. 4, pp. 811-815, Apr. 2020.

\bibitem{intro01-1} T. Huynh-The, C. Hua, J. Kim, S. Kim and D. Kim, ``Exploiting a low-cost CNN with skip connection for robust automatic modulation classification,'' in~\textit{Proc. 2020 IEEE Wireless Commun. Netw. Conf. (WCNC)}, Seoul, Korea (South), 2020, pp. 1-6.

\bibitem{intro01-2} Q.-V. Pham, N. T. Nguyen, T. Huynh-The, L. B. Le, K. Lee and W.-J. Hwang, ``Intelligent Radio Signal Processing: A Contemporary Survey,'' \textit{arXiv preprint arXiv:2008.08264}, 2020.

\bibitem{intro02} W. Wei and J. M. Mendel, ``Maximum-likelihood classification for digital amplitude-phase modulations,'' \textit{IEEE Trans. Commun.}, vol. 48, no. 2, pp. 189-193, Feb. 2000.

\bibitem{intro03} F. Hameed, O. A. Dobre and D. C. Popescu, ``On the likelihood-based approach to modulation classification,'' \textit{IEEE Trans. Wireless Commun.}, vol. 8, no. 12, pp. 5884-5892, Dec. 2009.

\bibitem{intro04} L. Han, F. Gao, Z. Li and O. A. Dobre, ``Low Complexity Automatic Modulation Classification Based on Order-Statistics,'' \textit{IEEE Trans. Wireless Commun.}, vol. 16, no. 1, pp. 400-411, Jan. 2017.

\bibitem{intro05} M. Abdelbar, W. H. Tranter and T. Bose, ``Cooperative Cumulants-Based Modulation Classification in Distributed Networks,'' \textit{IEEE Trans. on Cogn. Commun. Netw.}, vol. 4, no. 3, pp. 446-461, Sept. 2018.

\bibitem{intro06} Y. LeCun, Y. Bengio and G. Hinton, ``Deep Learning,'' \textit{Nature}, vol. 521, no. 7553, pp. 436–444, May 2015.

\bibitem{intro07} T. Huynh-The, C. Hua and D. Kim, ``Encoding Pose Features to Images with Data Augmentation for 3D Action Recognition,'' \textit{IEEE Trans. Ind. Informat.}, vol. 15, no. 5, pp. 3100-3111, May 2020.

\bibitem{intro07-1} T. Huynh-The, C. Hua, T. Ngo and D. Kim, ``Image representation of pose-transition feature for 3D skeleton-based action recognition,'' \textit{Inf. Sci.}, vol. 513, pp. 112-126, Mar. 2020.

\bibitem{intro07-2} T. Huynh-The, C. Hua, N. A. Tu and D. Kim, ``Learning 3D spatiotemporal gait feature by convolutional network for person identification,'' \textit{Neurocomputing}, vol. 397, pp. 192-202, July 2020.

\bibitem{intro08} Cam-Hao Hua \textit{et al.}, ``Bimodal learning via trilogy of skip-connection deep networks for diabetic retinopathy risk progression identification,'' \textit{Int. J. Med. Inform.}, vol. 132, pp. 103926, Dec. 2019.

\bibitem{intro08-1} T. Huynh-The, C. Hua, N. A. Tu and D. Kim, ``Physical Activity Recognition with Statistical-Deep Fusion Model using Multiple Sensory Data for Smart Health,'' \textit{IEEE Internet Things J.}, 2020, doi: 10.1109/JIOT.2020.3013272, in press.

\bibitem{intro09} T. J. O’Shea, T. Roy and T. C. Clancy, ``Over-the-Air Deep Learning Based Radio Signal Classification,'' \textit{IEEE J. Sel. Topics Signal Process.}, vol. 12, no. 1, pp. 168-179, Feb. 2018. 

\bibitem{work01} W. Xie, S. Hu, C. Yu, P. Zhu, X. Peng and J. Ouyang, ``Deep Learning in Digital Modulation Recognition Using High Order Cumulants,'' \textit{IEEE Access}, vol. 7, pp. 63760-63766, 2019.

\bibitem{work02} S. Hu, Y. Pei, P. P. Liang and Y. Liang, ``Deep Neural Network for Robust Modulation Classification Under Uncertain Noise Conditions,'' \textit{IEEE Trans. Veh. Technol.}, vol. 69, no. 1, pp. 564-577, Jan. 2020.

\bibitem{work03} F. Meng, P. Chen, L. Wu and X. Wang, ``Automatic Modulation Classification: A Deep Learning Enabled Approach,'' \textit{IEEE Trans. Veh. Technol.}, vol. 67, no. 11, pp. 10760-10772, Nov. 2018.

\bibitem{work04} Y. Zeng, M. Zhang, F. Han, Y. Gong and J. Zhang, ``Spectrum Analysis and Convolutional Neural Network for Automatic Modulation Recognition,'' \textit{IEEE Wireless Commun. Lett.}, vol. 8, no. 3, pp. 929-932, Jun. 2019.

\bibitem{work05} S. Huang \textit{et al.}, ``Automatic Modulation Classification Using Compressive Convolutional Neural Network,'' \textit{IEEE Access}, vol. 7, pp. 79636-79643, 2019.

\bibitem{work06} Y. Wang, M. Liu, J. Yang and G. Gui, ``Data-Driven Deep Learning for Automatic Modulation Recognition in Cognitive Radios,'' \textit{IEEE Trans. Veh. Technol.}, vol. 68, no. 4, pp. 4074-4077, Apr. 2019.

\bibitem{EPA01} 3GPP TS 36.101, ``Evolved Universal Terrestrial Radio Access (E-UTRA); User Equipment (UE) Radio Transmission and Reception,'' \textit{3rd Generation Partnership Project; Technical Specification Group Radio Access Network}.

\bibitem{EPA02} 3GPP TS 36.104 ``Evolved Universal Terrestrial Radio Access (E-UTRA); Base Station (BS) Radio Transmission and Reception,'' \textit{3rd Generation Partnership Project; Technical Specification Group Radio Access Network}.

\bibitem{method01}
C. Szegedy, V. Vanhoucke, S. Ioffe, J. Shlens and Z. Wojna, ``Rethinking the Inception Architecture for Computer Vision,'' in~\textit{Proc. 2016 IEEE Conf. Comput. Vis. Pattern Recognit. (CVPR)}, Las Vegas, NV, 2016, pp. 2818-2826.


\end{thebibliography}
\end{document}